\documentstyle[11pt,epsfig]{article}
\begin{document}
\vspace*{-1.8cm}
\begin{flushright}
\flushright{\bf LAL 99-55}\\
\vspace*{-0.5cm}
\flushright{October 1999}
\end{flushright}
\vskip 2.5 cm

\centerline {\LARGE\bf Probing Galactic structure using}
\vspace{2mm}
\centerline {\LARGE\bf micro-lensing with EROS-2}

\vskip 1.5 cm
\begin{center}
{\bf\Large   R. Ansari}\\
EROS collaboration
\end{center}

\begin{center}
{\bf\large Laboratoire de l'Acc\'el\'erateur Lin\'eaire}\\
{IN2P3-CNRS et Universit\'e de Paris-Sud, BP 34, F-91898 Orsay Cedex}
\end{center}
\vspace*{0.5cm}

\begin{abstract}
EROS has been monitoring few million stars in the Magellanic clouds,
as well as toward the Galactic bulge and spiral arms since 1996,
to search for microlensing events.
In this paper, we present briefly the EROS setup and scientific program 
and discuss the results obtained from our observations in four directions
in the Galactic plane, away from the bulge.
Seven light curves, out of 9.1 million stars observed in these
directions show luminosity variations interpreted as due to microlensing.
The averaged estimated optical depth 
$\bar{\tau} = 0.45^{+0.24}_{-0.11}$
is compatible with expectations from simple Galactic models.
Nonetheless a small excess of short time-scale events may be present
in the direction closest to the Galactic center.
\end{abstract}
\vspace*{0.5cm}

\section{Introduction}
Following Paczynski's 1986 proposal, several groups 
(EROS, MACHO, OGLE) started
systematic photometric surveys in the  beginning of the '90s
to search for microlensing events due compact objects (stars, brown dwarfs)
present in the Galactic disk, and possibly in the halo.

Since the first discoveries \cite{REF1,REF2,REF3}, hundreds
of events have been detected in the Galactic bulge direction, while only a few 
events ($\simeq 10-20$) have been observed toward the Magellanic clouds.

The event rates observed toward the Magellanic clouds (LMC, SMC) 
are significantly smaller than the expected rate from a standard halo, made 
of brown dwarves (MACHO's). On the contrary, the large number of events
seen in the direction of the bulge \cite{REF4,REF5}
have led to the hypothesis of a barred structure for the Galaxy.
This possibility, suggested by de Vaucouleurs in 1964, is
now supported by different observations: bulge micro-lensing rate,
photometric measurements \cite{REF6}, gas and star 
kinematics \cite{REF7,REF8} and star 
counting \cite{REF9}.

In order to disentangle contributions from different Galactic components 
(disk, bar, halo), EROS decided to design and build a second generation 
apparatus in 1994. We briefly describe this setup and our scientific
program in the next section.

\section{EROS-2 experimental setup and observation program}
The MARLY telescope (D=1 m, f/5) has been specially refurbished and 
fully automated for the EROS-2 survey (Bauer et al. 1997). The telescope
optics allows simultaneous imaging in two wide pass-bands 
$V_{Eros}$ ($\lambda_{peak} \simeq$ 560 nm), 
$R_{Eros}$ ($\lambda_{peak} \simeq$ 760 nm) 
over a one-square-degree field of vue. The two focal planes are 
equipped with CCD cameras, each made of a mosaic of 
8 $(2 \times 4)$ Loral $2048 \times 2048$ thick CCD's, covering 
a total field of $0.7^{\circ}$ (right ascension) 
$\times$ $1.4^{\circ}$ (declination). The pixel size is 0.6 arcsec, with 
a typical global seeing of 2 arcsec FWHM.
  
This telescope-camera system has been in operation at the European 
Southern Observatory in La Silla, Chile, since June 1996. 
The observations are carried out for three main scientific
programs. Type I supernovae are being used as standard
candles to probe the Universe geometry. EROS participate
in this effort through a powerful 
automated supernovae search ($z \sim 0.1$).
Around 60 SN have been discovered in the period 1997-1999.
We were able to determine the type for $\sim$ 25 SN, of 
which 20 have been identified as type Ia supernovae.
The mean discovery rate is around 1 SN / 2 hours observing time
($\sim 1 SN / 10-20 deg^2$).

A small fraction of the EROS telescope time is used  
to search for Red dwarves through proper motion detection, while
most of the telescope time is devoted to photometric surveys for
microlensing searches.
Systematic photometric surveys are being carried out in several directions,
with typical sampling times of a few days. 
Around 80 fields are being monitored toward the Large Magellanic Cloud (LMC)
and 10 fields toward the SMC. A large area in direction of the 
bulge is also included in our survey (CG , $\sim$ 150 fields) as well 
as 29 fields in the Galactic plane, away from the bulge.

\section{EROS-2 GSA microlensing search}
The 29 Galactic plane fields (GSA) are grouped in four directions 
$(\beta_{Sct} , \gamma_{Sct}$, $\gamma_{Nor} , \theta_{Mus} )$ 
and cover a wide range of longitude.
The three year data set discussed here contains 9 million 
light curves : 2.1 towards $\beta_{Sct}$, 1.8 towards $\gamma_{Sct}$, 
3.0 towards $\gamma_{Nor}$ and 2.1 towards $\theta_{Mus}$.
The observation period for these fields span July 1996 - November 1998, 
except for $\theta_{Mus}$ which has been monitored since January 1997.

The first steps of the event selection criteria require the presence 
of a single bump, simultaneous in two colors. To reject variable stars,
We require stability of the light curve outside the bump 
and compatibility with the expected microlensing shape 
(cuts on $\chi^2$ from microlensing fit). 

Detailed description of EROS GSA data set and event selection can 
be found in \cite{REF10,REF11}. 

Seven light curves satisfy all the selection criteria and are 
labeled GSA1 to 7. The characteristics for these seven (7) 
microlensing events can be found in table \ref{caract}. 

GSA1 is a large magnification event, while GSA2 exhibit periodic
luminosity modulation during amplification. Figure 1a
shows the GSA2 light curve which is interpreted as a microlensing
effect on a binary source system.

\begin{table*}[htbp]
\caption[]{Characteristics of the 7 microlensing candidates and contribution to the optical depth.}
\begin{center}
{\small 
\begin{tabular}{lllll}
\hline\noalign{\smallskip}
Candidate & GSA1 & GSA2 & GSA3 & GSA4 \\
\noalign{\smallskip}
\hline\noalign{\smallskip}
field & $\gamma\; Sct $ & $\gamma\; Nor$ & $\gamma\; Nor$  & $\gamma\; Sct$ \\
$\alpha$(h:m:s)	& 18:29:09.0	& 16:11:50.2	& 16:16:26.7 & 18:32:26.0 \\
$\delta$(d:m:s)	& -14:15:09     & -52:56:49     & -54:37:49  & -12:56:04  \\
$R_{EROS}$ - $V_{EROS}$	& 17.7 - 20.7	& 17.8 - 19.4 	& 17.5 - 18.6 & 17.1 - 17.9 \\

$\Delta t = R_E/V_T$ (days) & $73.5\pm 1.4$ & $98.3\pm 0.9$ & $70.0\pm 2.0$ & $23.9\pm 1.1$\\
Max. magnification	& $26.5\pm 0.6$	& $3.05\pm 0.02$ & $1.89\pm 0.01$ & $1.72\pm 0.02$ \\

$\chi^2$ of best fit	& 185.7/163 	& 551/425 	& 445/427  & 337/195 \\
contribution to $\tau$ ($\times 10^6$) & 0.51 & 0.15 & 0.12 & 0.30 \\
\noalign{\smallskip}
\end{tabular}
\begin{tabular}{llll}
\hline\noalign{\smallskip}
Candidate & GSA5 & GSA6 & GSA7 \\
\noalign{\smallskip}
\hline\noalign{\smallskip}
field &  $\gamma\; Sct$ &  $\gamma\; Sct$ &  $\gamma\; Sct$  \\
$\alpha$(h:m:s)	& 18:32:12.0 & 18:33:56.7 & 18:34:10.0	\\
$\delta$(d:m:s)	& -12:55:16 & -14:33:52   & -14:03:40   \\
$R_{EROS}$ - $V_{EROS}$	  & 17.9 - 19.9		& 17.2 - 18.5 & 17.5 - 18.7 \\
$\Delta t = R_E/V_T$ (days) & $59.0\pm 5.5$	& $37.9\pm 5.0$	& $6.20 \pm 0.50$ \\
Max. magnification	& $1.71\pm 0.03$ & $1.35\pm 0.02$ & $2.70 \pm 0.30 $	\\
$\chi^2$ of best fit	& 122/186 	& 104/170 &    87 / 181	\\							
contribution to $\tau$ ($\times 10^6$) & 0.44 & 0.35 & 0.22 \\
\noalign{\smallskip}
\hline
\end{tabular}
}
\end{center}
\label{caract}
\end{table*}
 
The efficiency of the selection process has been computed using
montecarlo generated light curves, where randomly generated microlensing
effect where superimposed on a representative sample of observed 
light curves. 

We have computed the expected optical depths using a three component
model (bar, disk, halo) to represent deflector distribution in the Galaxy.
The distance distribution of the source stars is poorly known. 
Different studies indicate distances in the range 5-10 kpc 
\cite{REF12,REF10}. An average distance of $\sim 7$ kpc
has been used in this paper. Figure 1b shows the 
expected optical depth as a function of the Galactic longitude,
for two sets of bar parameters at b = 2$^{\circ}$.5. 
For a given target an estimation of the optical depth (or a limit) 
can be computed using the expression:
$$ \tau = \pi / 2 \times 1 / ( N_{obs}T_{obs} ) \sum_{events}
\Delta t / \epsilon (\Delta t ) $$
where $N_{obs}$ is the number of monitored stars and
$T_{obs}$ the duration of the search period.
The corresponding values for the four targets are also indicated on
figure 1.



\section{Conclusion}
We find an estimated optical depth averaged over the four directions  
$$ \bar\tau = 0.45^{+0.24}_{-0.11} \times 10^{-6} $$
in agreement with expectations. However, as shown on figure 1,
optical depths or limits computed for each direction indicates a small
excess of events toward $\gamma_{Sct}$ direction, compared to other
directions.

Provided that this is not due to a statistical fluctuation, the
shorter event time scale toward $\gamma_{Sct}$ favors an explanation 
based on an increased contribution from the bar.
A significant difference in the source stars distance between different
targets is another plausible explanation for the observed excess, 
$\gamma_{Sct}$ stars may be located at a larger distance, compared
to other directions.

A larger number of events, with independent source stars distance 
determination will help in providing a more reliable explanation.

\vspace*{-0.05cm}
\begin{figure}
\begin{center}
{\bf a)}\epsfig{file=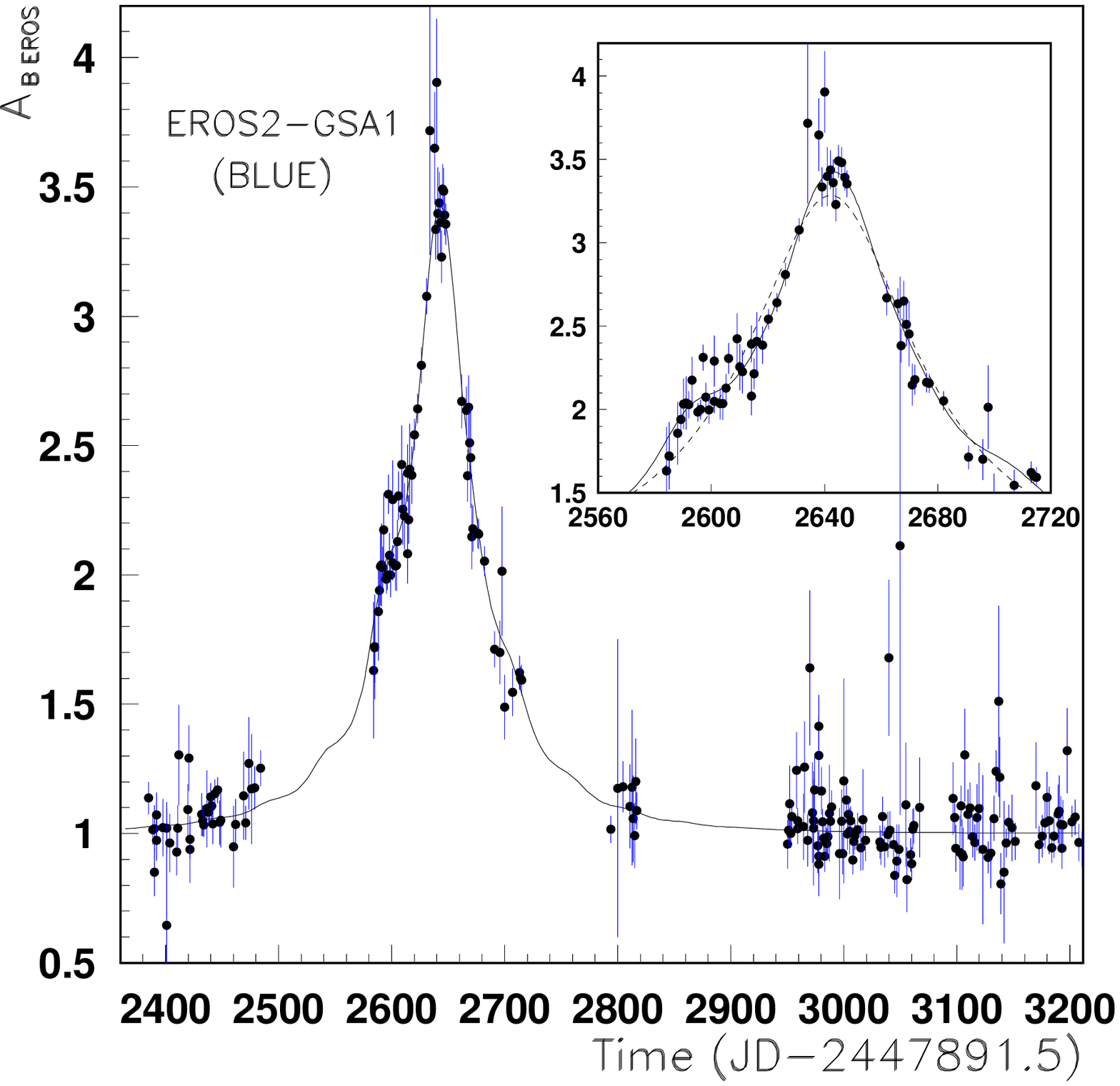,width=10cm}
\end{center}
\vspace*{-0.3cm}
\begin{center}
{\bf b)}\epsfig{file=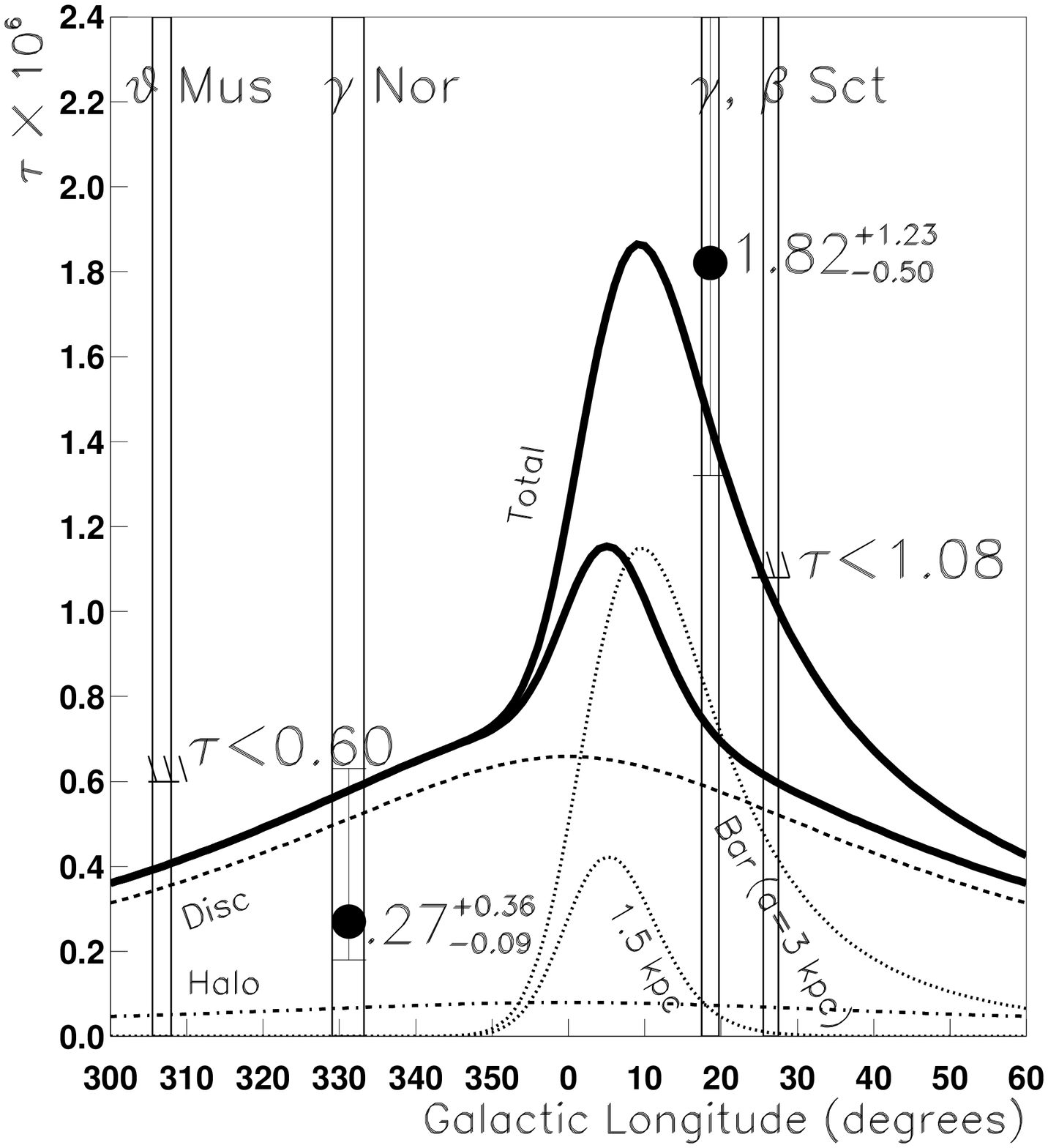,width=10cm}
\end{center}
\caption{a: Light curve for candidate GSA2 ($V_{Eros}$), b: 
Expected optical depth ($\times\ 10^6$) up to 7 kpc for different components of the Milky Way. Estimated optical depths (or upper limits) for the four monitored directions are also shown.}
\label{myfig}
\end{figure}

\end{document}